\documentclass[final,5p,times]{elsarticle}

\usepackage[linktocpage=true,unicode=true,ps2pdf]{hyperref}
\usepackage{hypcap}
\usepackage{amsmath,amssymb,amsfonts}
\usepackage{graphicx}
\usepackage{color}

\graphicspath{{./figs/}}
\begin{document}

\begin{frontmatter}

\title{Galactic positrons and electrons from dark matter and astrophysical sources.}
\author{Roberto Lineros}
\ead{lineros@to.infn.it}
\address{Dipartimento di Fisica Teorica, Universit\`{a} di Torino
and INFN -- Sezione Torino, via P. Giuria 1, I--10125 Torino, Italy.}

\begin{abstract}
The electron and positron cosmic rays observations have impulsed a hot debate regarding the origin of such particles.
Their propagation in the galactic medium is modeled according to a successfully tested two--zone propagation model.
The theoretical uncertainties related to their propagation and production are studied for three cases: secondary production, Dark Matter annihilation, and originating from supernovae remnants.\\ 
\end{abstract}


\end{frontmatter}

%\begin{flushleft}
%  Preprint DFTT ?? and LAPTH ??
%\end{flushleft}

\section{Introduction}
\label{sec1}

The latest experimental results: the \emph{positron fraction}~\cite{DuVernois:2001ApJ, 2009Natur.458..607A} and the \emph{electrons+positrons flux}~\cite{Chang:2008zz, 2009PhRvL.102r1101A}, have presented an interesting deviation in respect with expected signals for energies above $\sim$20 GeV.\\

In this work, we describe the propagation model used to study electron and positron cosmic rays (CR), which has been used to describe successfully Nuclei CR (NCR) observations. 
We discuss the theoretical uncertainties paying attention to propagation, that plays a very important role for the study of the electron/positron signal.
Moreover, we study different mechanisms to produce electrons and positrons: secondary production, annihilating dark matter (DM), and supernova (SN) remnants. In each of them, we discuss uncertainties and its contribution to the signal.\\

\section{Propagation model}
\label{sec2}

The CR propagation is modeled using a two--zone propagation model~\cite{1980Ap&SS..68..295G}, in which, CR number density per unit of energy $(\psi)$ is governed by the following transport equation~\cite{1980Ap&SS..68..295G}:
\begin{equation}
\label{eq:te}
\frac{\partial \psi }{\partial t} + \nabla \cdot J_x + \frac{\partial J_\epsilon}{\partial \epsilon} = q_{\textnormal{src}} \, .
\end{equation}
The CR propagation takes place inside a cylinder -- centered in the Galactic Center and oriented like the Galactic Plane -- with a radius set in 20~kpc and half--thickness $L_z$ constrained by NCR observations~\cite{2001ApJ...555..585M}, typically it varies in the range of 1--20~kpc \\

The first term in equation~\ref{eq:te} is related to the evolution in time. 
The second one depends on $J_x$, 
\begin{equation}
\label{eq:jx}
  J_x = - K_0 \epsilon^{\delta} \, \nabla \psi + \bf{V}_c \, \psi \, ,
\end{equation}
which models the spatial diffusion due to magnetic inhomogeneities, the diffusion term is assumed to be a power law in energy expressed in terms of the two parameters, $K_0$ and $\delta$. 
As well, $J_x$ takes into account advection and convection produced by the Galactic Wind $\bf{V}_c$.\\

The CR evolution in energy is mainly modelled by the third term in equation~\ref{eq:te}, where 
\begin{equation}
  \label{eq:en_ev}
J_\epsilon = - \frac{dE}{dt} \, \psi - \frac{\nabla \cdot \bf{V}_c}{3} \epsilon \, \psi  + K_{\epsilon \epsilon} \frac{2}{\epsilon} \psi - K_{\epsilon \epsilon}  \frac{\partial}{\partial \epsilon} \psi \, .
\end{equation}

In equation~\ref{eq:en_ev}, $\displaystyle\frac{dE}{dt}$ corresponds to the energy--loss term, which contains effects as: Inverse Compton scattering (ICS) with radiation fields, synchrotron radiation, ionization of the interstellar medium (ISM), among others~\cite{1998ApJ...493..694M,2008arXiv0812.4272L}.
Adiabatic losses produced by the Galactic wind are also present.
The last two terms are related to CR reacceleration. In the standard theory of reacceleration the coefficient $K_{\epsilon\epsilon}$ is expressed in terms of the spatial diffusion parameters (Equation~\ref{eq:jx}) and the \emph{\'{A}lfven velocity} $V_a$ (see~\cite{Maurin:2002ua}):
\begin{equation}
K_{\epsilon \epsilon} = \frac{2}{9} \frac{V_a^2}{K_0} \epsilon^{2-\delta} \, .
\end{equation}

For ultrarelativistic electrons and positrons, ICS with the CMB and starlight, and Synchrotron radiation induced by galactic magnetic fields rule the energy evolution.
For energies below 3--5 GeV, the Thomson limit for ICS is a safe approximation. However, for energies above the former limit, this approximation is no longer valid and it is necessary to use the Klein--Nishina approach~\cite{2009prep, 1970RvMP...42..237B}.\\
In equation~\ref{eq:te}, $q_{\textnormal{src}}$ is the source term that describes how, when and where CR are injected into the propagation zone. 
Therefore, it depends on the type of source which is studied. 
The source term is described in more details in section~\ref{sec3} for three of the most interesting sources of electrons and positrons.\\
\subsection{Propagation uncertainties}

% mini intro
The global picture of how CR propagate in the Galaxy is far to be completely understood.
Nevertheless, one of the requirements of a propagation model is to reproduce observations of many CR species.
The analysis of the ratio boron/carbon (B/C) is sensitive to different propagation models.
Indeed boron is mainly produced by spallation of NCR on the ISM and is not accelerated in SN remnants.\\
%instead of stellar activity.\\
%

%
For the study of electron and positron CR, we use all B/C compatible parameter sets~\cite{2001ApJ...555..585M}.
Among them we highlight three sets: MAX, MED, and MIN (Table~\ref{table:prop}).
The set MAX (MIN) maximizes (minimizes) the B/C ratio and the set MED corresponds to the B/C best fit.\\

%
%For the study of electron and positron CR, we use the B/C compatible volume~\cite{2001ApJ...555..585M}. 
%
%Inside this, we denote three special sets: MAX, MED, and MIN (Table~\ref{table:prop}). The set MAX (MIN) maximizes (minimizes) the B/C ratio and the set MED corresponds to the B/C best fit.\\

\begin{table}[t]
\centering
\resizebox{0.9\hsize}{!}{\begin{tabular}{|c||c|c|c|c|c|c|}
\hline
Model  & $\delta$ & $K_0$ & $L_z$ & $V_{c}$ & $V_{a}$ \\
       &          & [kpc$^2$/Myr] & [kpc] & [km/s] & [km/s] \\
\hline
MIN  & 0.85 &  0.0016 & 1  & 13.5 &  22.4 \\
MED  & 0.70 &  0.0112 & 4  & 12   &  52.9 \\
MAX  & 0.46 &  0.0765 & 15 &  5   & 117.6 \\
\hline
\end{tabular}
}
\caption{
\label{table:prop}
Typical combinations of diffusion parameters that are compatible with the B/C
analysis~\cite{2001ApJ...555..585M}. 
}
\end{table}

\section{Electron and positron production}
\label{sec3}

The source of electrons and positrons in the Galaxy is not unique.
In fact, various mechanisms have been suggested to explain the observations.
First of all, we distinguish the ones with an astrophysical origin like SN remnants~\cite{2009prep} and Pulsars~\cite{2008arXiv0812.4457P}.
There are also secondary production i.e. interaction of CR with the ISM. 
And finally, the exotic component like DM.
In this section, we describe most of them.

%
%
%%%%%%%%%%%%%%%%%%%%%%%%%%%%%%%%%%%%%%%%%%%%%%%%%%%%%%%%%%%%%%%%%%%%%%%%%%%%%%%%%%%%%%%%%%%%%
\subsection{Secondary production}
% basic introduction
The interaction of NCR with the ISM, mainly composed by hydrogen and helium, leads the secondary production of electrons and positrons.\\

The secondary production is conditioned by three principal factors: the distribution and energy spectra of NCR, ISM gas distribution, and nuclear cross sections.
For instance, the term due to interaction between proton CR and ISM hydrogen is:
\begin{equation}
  \label{eq:sec_src}
  q_{\textnormal{sec}}({\bf x},\epsilon) = 4\pi \, n_{\textnormal{H}}({\bf x}) \int \, d E_p \Phi_p({\bf x},E_p) \, \frac{d \sigma_{p\textnormal{H}}}{d \epsilon}(E_p,\epsilon) \, ,
\end{equation}
where $n_{\textnormal{H}}({\bf x})$ is the hydrogen number density, $\Phi_p({\bf x},E_p)$ is the proton flux and $\displaystyle\frac{d \sigma_{p\textnormal{H}}}{d \epsilon}$ is the inclusive production cross section of electron or positrons~\cite{2008arXiv0809.5268D,1998ApJ...493..694M}.\\

Quite large theoretical uncertainties related to secondary production come from nuclear cross sections and NCR fluxes.
However, the biggest uncertainties remain the ones related to propagation parameters~\cite{2008arXiv0809.5268D}.

%
%
%
%%%%%%%%%%%%%%%%%%%%%%%%%%%%%%%%%%%%%%%%%%%%%%%%%%%%%%%%%%%%%%%%%%%%%%%%%%%%%%%%%%%%%%%%%%%%%
\subsection{Dark Matter annihilation}
Invoquing DM annihilation could be an explaination for the \emph{positron fraction} and the \emph{electron+positron flux} features.
According to N-body simulation and structure formation models, galaxies like the Milky Way are supposed to be embedded into DM haloes which are denser in the surrounding of galactic center.
Over-densities increase the DM annihilation rate and would enhance the DM component in the CR signal.\\
The annihilation mechanism depends on the particle physics theory behind DM, but in general terms, it is depicted as:
\begin{equation}
\rm{DM} + \overline{\rm{DM}} \longrightarrow F + F^{'} \rightarrow e^{\pm} + X \, ,
\end{equation}
where $F$ and $F^{'}$ are particles produced directly by the annihilation.
By hadronization and/or decay they will then produce  electrons, positrons, and other particles.\\
In this case, the source term is:
\begin{equation}
  \label{eq:dm_st}
  q_{\rm{DM}}({\bf x},\epsilon) = \alpha \, \langle \sigma_{\textnormal{ann}} v \rangle \, \frac{\rho^2({\bf x})}{m^2_{\chi}} \, \frac{dn}{d\epsilon}(\epsilon)
\end{equation}
where $\alpha$ is a factor that depends on whether the DM particle is its own antiparticle ($\alpha = 1/4$) or not ($\alpha = 1/2$). $m_{\chi}$ is the DM particle mass. 
In equation~\ref{eq:dm_st}, we also distinguish three other terms:
\begin{itemize}
\item $\langle \sigma_{\textnormal{ann}} v \rangle$ is the thermally averaged annihilation cross section which depends of the particle physics model.\\
\item $\rho^2(\bf{x})$ corresponds to the annihilation distribution which depends directly on the DM distribution. The most common DM distributions are: Cored Isothermal, Navarro-Frenk-White~\cite{1997ApJ...490..493N} and Moore~\cite{2004MNRAS.353..624D}.
\item $\displaystyle\frac{dn}{d\epsilon}$ is the multiplicity distribution per single annihilation event which comes from hadronization of quarks and decay of particles.
\end{itemize}

%%%%%%%%%%%%%%%%%%%%%%%%%%%%%%%%%%%%%%%%%%%%%%%%%%%%%%%%%%%%%%%%%%%%%%%%%%%%%%%%%%%%%%%%%%%%%
\subsection{Electrons from Supernovae}

SN are expected to be one of the principal sources of galactic primary CR.
Shock--waves originated from the explosion provide a good mechanism to (re)accelerate charged particles to very high energy.
Hence, SN are good candidates to explain the primary component present in the electron signal~\cite{2009prep}.
However, current primary electron CR models are so simple that it becomes hard to make predictions.\\

A new re-estimation of the primary component is performed by taking in consideration some natural aspects~\cite{2009prep} like: inhomogeneities in the distribution of nearby SN, injection electron spectra, and time dependence propagation, among others.

\section{Discussion}
\label{sec4}

\begin{figure*}[t]
\centering
\resizebox{0.9\hsize}{!}{\includegraphics[width=100pt, angle=270]{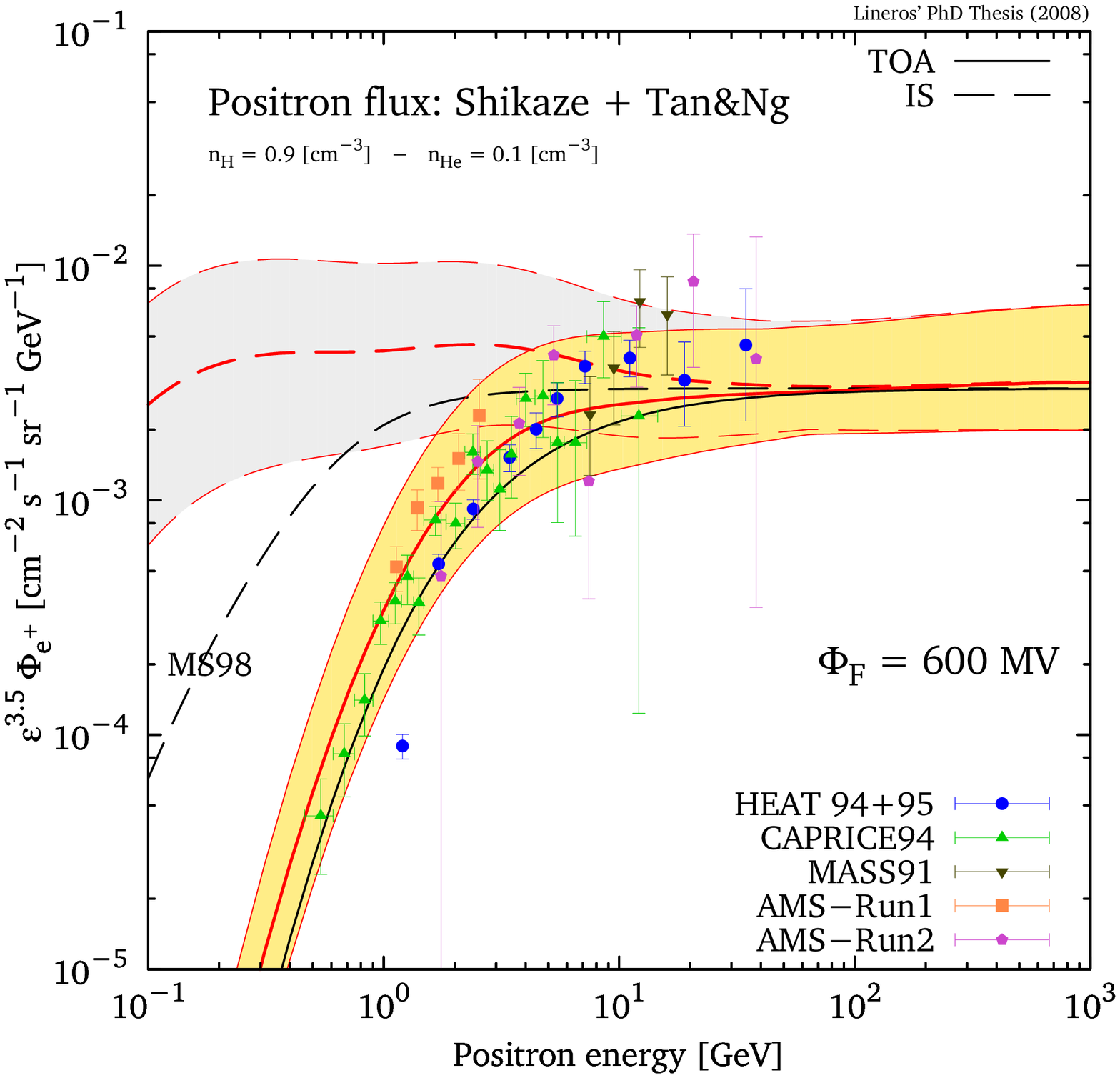}
\includegraphics[width=100pt, angle=270]{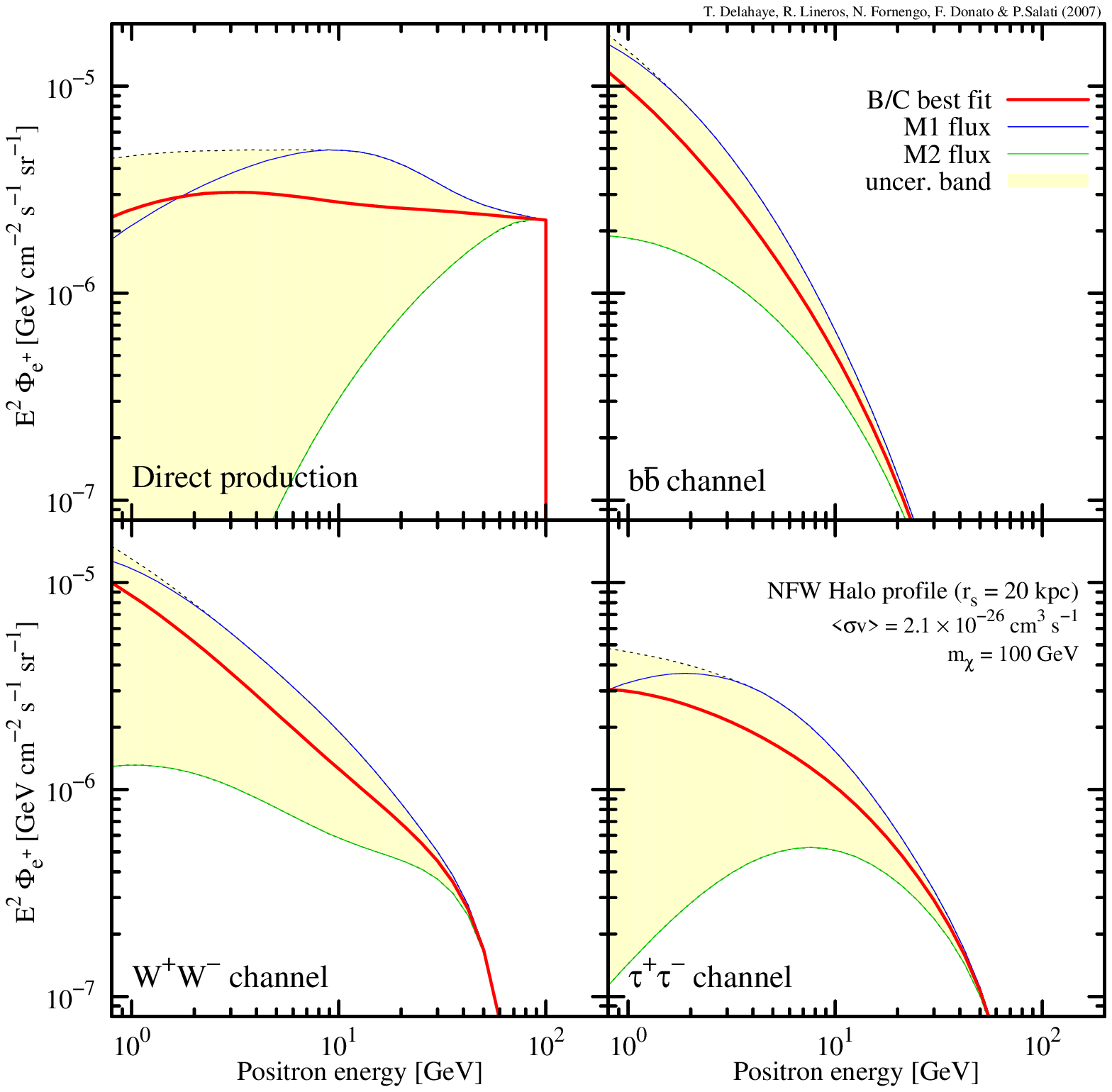}}
\caption{
\label{f:teo_unc} Flux of secondary positrons (left) and positron from DM annihilation (right) versus energy.
In both cases, propagation uncertainties from B/C analysis~\cite{2001ApJ...555..585M} are taken into account.
% talk about secondary plot
In the case of secondary positrons, our calculations are encompass with current available data \cite{2008arXiv0809.5268D, 2008arXiv0812.4272L,2009arXiv0905.2144D}.
% talk about dm
In the case of DM annihilation, the shape of the positron flux mainly depends on the annihilation channel and the chosen propagation set~\cite{2008PhRvD..77f3527D,2008ICRC....4..705F}. Direct annihilation into electrons and positrons produces a harder spectra than cases where DM annihilates into pairs of quarks or gauge bosons. 
}
\end{figure*}

% intro to fig 1
The propagation uncertainties have a big impact on secondary positron flux (Figure~\ref{f:teo_unc}).
The variety of B/C compatible propagation parameter sets results in a band.
This band shows that the propagation uncertainty encompasses the experimental positron flux.\\
In the case of DM (Figure~\ref{f:teo_unc}), we observe that the annihilation spectra produce different propagated fluxes.
The propagation uncertainties produce different behaviours in the flux.
And all of them have a non negligible effect in the positron fraction~\cite{2008PhRvD..77f3527D}. \\

% fig1

\begin{figure*}[t]
\centering
\resizebox{0.9\hsize}{!}{\includegraphics[angle=270,width=100pt]{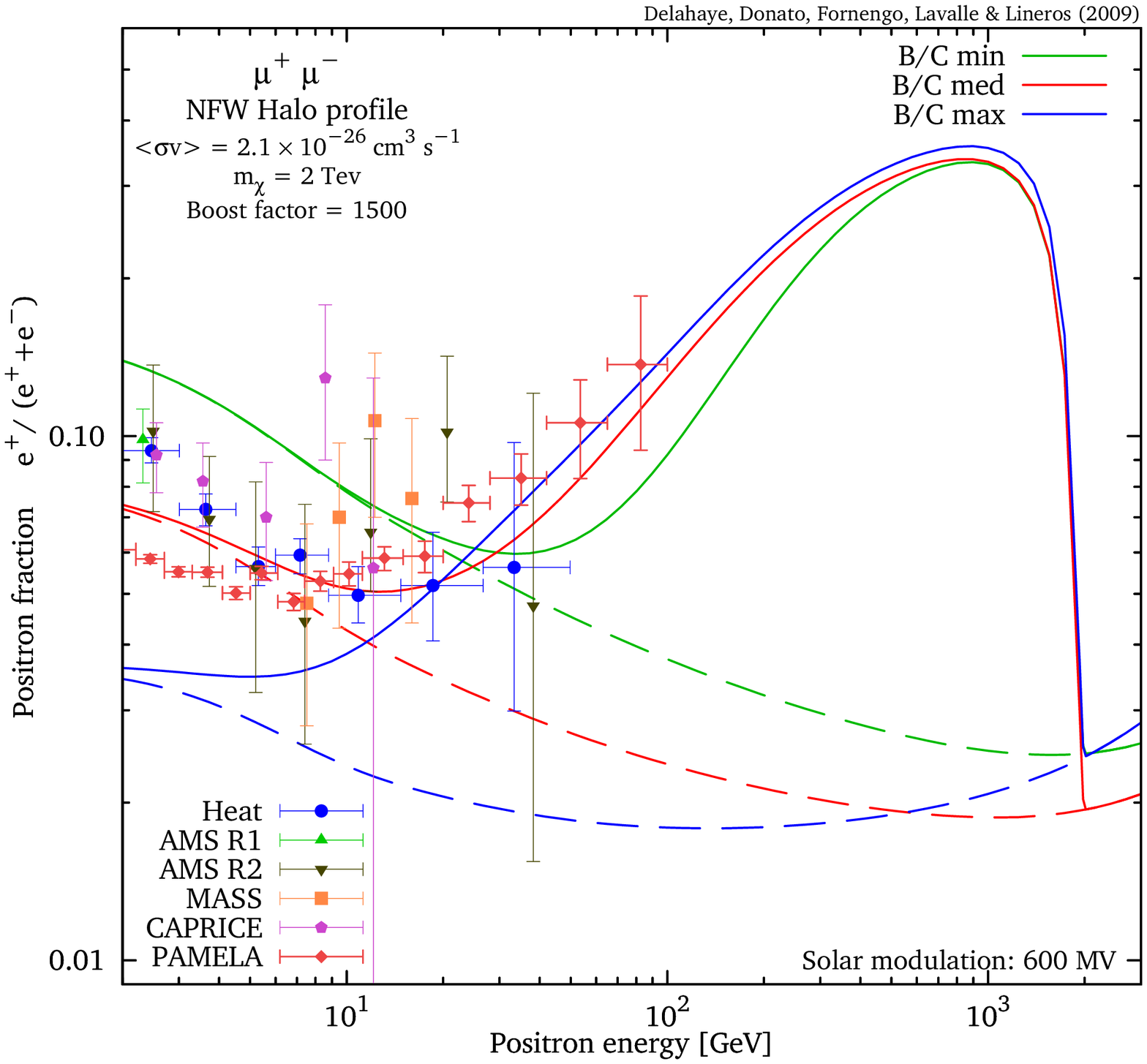}
\includegraphics[angle=270,width=100pt]{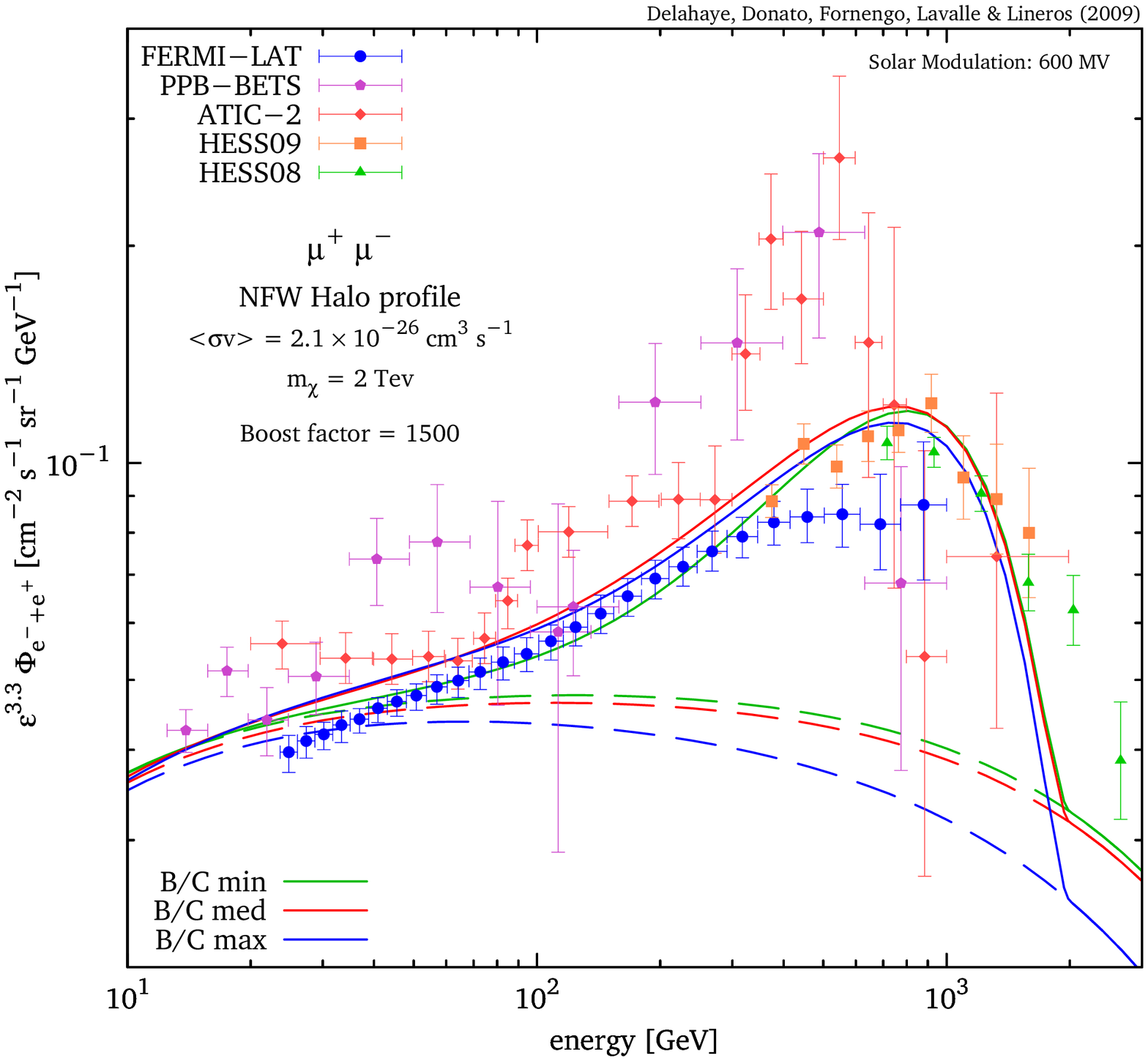}}
\caption{
\label{f:dm_sol} 
% naive description
Positron fraction (left) and total electron+positron flux (right) versus energy.
% common description 
A possible explanation of the \emph{positron fraction} and the \emph{electron+positron} flux is given in terms of annihilating DM.
% Details about DM
For instance, we present the case of a 2--TeV DM particle which annihilates directly into muons. 
Solutions for different B/C compatible propagation sets~\cite{2001ApJ...555..585M} are encompass with current experimental data. 
Nevertheless, it requires a large value of boost factor which seems to be hardly explained in standard DM models compatible with other CR observation~\cite{2009arXiv0905.0480M}.
Current available data from HEAT~\cite{DuVernois:2001ApJ}, AMS~\cite{Alcaraz:2000PhLB}, MASS~\cite{Grimani:2002}, CAPRICE~\cite{Boezio:2000}, PAMELA~\cite{2009Natur.458..607A}, FERMI-LAT~\cite{2009PhRvL.102r1101A}, PPB-BETS~\cite{Torii:2001,Torii:2008}, ATIC~\cite{Chang:2008zz} and HESS~\cite{2008PhRvL.101z1104A,2009arXiv0905.0105H} are also shown.
}
\end{figure*}

In figure~\ref{f:dm_sol}, we study the case of a 2--TeV DM particles that annihilate into muon pairs.
This kind of signal results compatible with current observations, but is not unique. 
On the other hand, the DM signal needs to be enhanced in a factor as larger as 1500, usually known as the \emph{boost factor problem}.
To reproduce actual experimental results, we require an extremely large enhancement. 
Most of DM models cannot naturally explain such big enhancement.
This problem motivates us to question and re-estimate the current standard background of electrons~\cite{2009prep}.\\

%fig2

\section{Conclusions}
\label{sec5}

The improvement in current electron and positron measurements and data has revealed a very interesting puzzle.
Different solution have been presented during the years.\\

We present the importance of theoretical uncertainties related to propagation and production of electron and positron CR~\cite{2008arXiv0812.4272L, 2008arXiv0809.5268D,2009arXiv0905.2144D,2008PhRvD..77f3527D}. 
We stress the necessity to re-estimate secondary and primary component, and to consider already known sources in order to discard/confirm possible presence of an undiscovered component, like the case of annihilating DM.\\

The near future is very exciting. Experiments as PAMELA, ATIC, FERMI and future AMS–02 will give valuable information to complete part of the electron/positron puzzle.

\section*{Acknowledgements}
\begin{small}
Work supported by research grants funded jointly by Ministero dell'Istruzione, dell'Universit\`a e della Ricerca (MIUR), by Universit\`a di Torino (UniTO), by Istituto Nazionale di Fisica Nucleare (INFN) within the Astroparticle Physics Project and by the Italian Space Agency (ASI) under contract N I/088/06/0.
R.L. acknowledges to T. Delahaye, F. Donato, N. Fornengo, J. Lavalle, P. Salati, and R. Taillet for unvaluable comments and suggestions.
\end{small}
\bibliographystyle{elsarticle-num}
\bibliography{refs}

\begin{thebibliography}{10}
\expandafter\ifx\csname url\endcsname\relax
  \def\url#1{\texttt{#1}}\fi
\expandafter\ifx\csname urlprefix\endcsname\relax\def\urlprefix{URL }\fi
\expandafter\ifx\csname href\endcsname\relax
  \def\href#1#2{#2} \def\path#1{#1}\fi

\bibitem{DuVernois:2001ApJ}
M.~A. {DuVernois}, et~al., ApJ 559 (2001) 296--303.
\newblock \href {http://dx.doi.org/10.1086/322324} {\path{doi:10.1086/322324}}.

\bibitem{2009Natur.458..607A}
O.~{Adriani}, et~al., Nature 458 (2009) 607--609.
\newblock \href {http://arxiv.org/abs/0810.4995} {\path{arXiv:0810.4995}},
  \href {http://dx.doi.org/10.1038/nature07942}
  {\path{doi:10.1038/nature07942}}.

\bibitem{Chang:2008zz}
J.~Chang, et~al., Nature 456 (2008) 362--365.
\newblock \href {http://dx.doi.org/10.1038/nature07477}
  {\path{doi:10.1038/nature07477}}.

\bibitem{2009PhRvL.102r1101A}
A.~A. {Abdo}, et~al., Physical Review Letters 102~(18) (2009) 181101--+.
\newblock \href {http://arxiv.org/abs/0905.0025} {\path{arXiv:0905.0025}},
  \href {http://dx.doi.org/10.1103/PhysRevLett.102.181101}
  {\path{doi:10.1103/PhysRevLett.102.181101}}.

\bibitem{1980Ap&SS..68..295G}
V.~L. {Ginzburg}, et~al., ApSS 68 (1980) 295--314.
\newblock \href {http://dx.doi.org/10.1007/BF00639701}
  {\path{doi:10.1007/BF00639701}}.

\bibitem{2001ApJ...555..585M}
D.~{Maurin}, et~al., ApJ 555 (2001) 585--596.
\newblock \href {http://arxiv.org/abs/arXiv:astro-ph/0101231}
  {\path{arXiv:arXiv:astro-ph/0101231}}, \href
  {http://dx.doi.org/10.1086/321496} {\path{doi:10.1086/321496}}.

\bibitem{1998ApJ...493..694M}
I.~V. {Moskalenko}, et~al., ApJ 493 (1998) 694--+.
\newblock \href {http://arxiv.org/abs/arXiv:astro-ph/9710124}
  {\path{arXiv:arXiv:astro-ph/9710124}}, \href
  {http://dx.doi.org/10.1086/305152} {\path{doi:10.1086/305152}}.

\bibitem{2008arXiv0812.4272L}
R.~A. {Lineros}, Ph.D. thesis, Universit\`a degli Studi di Torino, dipartimento
  di fisica (Dec. 2008).
\newblock \href {http://arxiv.org/abs/0812.4272} {\path{arXiv:0812.4272}}.

\bibitem{Maurin:2002ua}
D.~Maurin, et~al.\href {http://arxiv.org/abs/astro-ph/0212111}
  {\path{arXiv:astro-ph/0212111}}.

\bibitem{2009prep}
T.~{Delahaye}, J.~{Lavalle}, R.~{Lineros}, F.~{Donato}, N.~{Fornengo}, \emph{in
  preparation}, Preprint DFTT 51/2009 and LAPTH 1339/09.

\bibitem{1970RvMP...42..237B}
G.~R. {Blumenthal}, R.~J. {Gould}, Reviews of Modern Physics 42 (1970)
  237--271.
\newblock \href {http://dx.doi.org/10.1103/RevModPhys.42.237}
  {\path{doi:10.1103/RevModPhys.42.237}}.

\bibitem{2008arXiv0812.4457P}
S.~{Profumo}, ArXiv e-prints\href {http://arxiv.org/abs/0812.4457}
  {\path{arXiv:0812.4457}}.

\bibitem{2008arXiv0809.5268D}
T.~{Delahaye}, F.~{Donato}, N.~{Fornengo}, J.~{Lavalle}, R.~{Lineros},
  P.~{Salati}, R.~{Taillet}, ArXiv e-prints\href
  {http://arxiv.org/abs/0809.5268} {\path{arXiv:0809.5268}}.

\bibitem{1997ApJ...490..493N}
J.~F. {Navarro}, et~al., ApJ 490 (1997) 493--+.
\newblock \href {http://arxiv.org/abs/arXiv:astro-ph/9611107}
  {\path{arXiv:arXiv:astro-ph/9611107}}, \href
  {http://dx.doi.org/10.1086/304888} {\path{doi:10.1086/304888}}.

\bibitem{2004MNRAS.353..624D}
J.~{Diemand}, et~al., MNRAS 353 (2004) 624--632.
\newblock \href {http://arxiv.org/abs/arXiv:astro-ph/0402267}
  {\path{arXiv:arXiv:astro-ph/0402267}}, \href
  {http://dx.doi.org/10.1111/j.1365-2966.2004.08094.x}
  {\path{doi:10.1111/j.1365-2966.2004.08094.x}}.

\bibitem{2008PhRvD..77f3527D}
T.~{Delahaye}, R.~{Lineros}, F.~{Donato}, N.~{Fornengo}, P.~{Salati}, PRD
  77~(6) (2008) 063527--+.
\newblock \href {http://arxiv.org/abs/0712.2312} {\path{arXiv:0712.2312}},
  \href {http://dx.doi.org/10.1103/PhysRevD.77.063527}
  {\path{doi:10.1103/PhysRevD.77.063527}}.

\bibitem{2009arXiv0905.2144D}
T.~{Delahaye}, P.~{Brun}, F.~{Donato}, N.~{Fornengo}, J.~{Lavalle},
  R.~{Lineros}, R.~{Taillet}, P.~{Salati}, 2009.
\newblock \href {http://arxiv.org/abs/0905.2144} {\path{arXiv:0905.2144}}.

\bibitem{2008ICRC....4..705F}
N.~{Fornengo}, T.~{Delahaye}, R.~{Lineros}, {et al.}, in: International Cosmic
  Ray Conference, Vol.~4 of International Cosmic Ray Conference, 2008, pp.
  705--708.

\bibitem{2009arXiv0905.0480M}
P.~{Meade}, et~al., ArXiv e-prints\href {http://arxiv.org/abs/0905.0480}
  {\path{arXiv:0905.0480}}.

\bibitem{Alcaraz:2000PhLB}
J.~{Alcaraz}, et~al., Physics Letters B 484 (2000) 10--22.

\bibitem{Grimani:2002}
C.~{Grimani}, et~al., A\&A 392 (2002) 287--294.
\newblock \href {http://dx.doi.org/10.1051/0004-6361:20020845}
  {\path{doi:10.1051/0004-6361:20020845}}.

\bibitem{Boezio:2000}
M.~Boezio, et~al., Astrophys. J. 532 (2000) 653--669.
\newblock \href {http://dx.doi.org/10.1086/308545} {\path{doi:10.1086/308545}}.

\bibitem{Torii:2001}
S.~{Torii}, et~al., ApJ 559 (2001) 973--984.
\newblock \href {http://dx.doi.org/10.1086/322274} {\path{doi:10.1086/322274}}.

\bibitem{Torii:2008}
S.~{Torii}, et~al., ArXiv e-prints\href {http://arxiv.org/abs/0809.0760}
  {\path{arXiv:0809.0760}}.

\bibitem{2008PhRvL.101z1104A}
F.~{Aharonian}, et~al., Physical Review Letters 101~(26) (2008) 261104--+.
\newblock \href {http://arxiv.org/abs/0811.3894} {\path{arXiv:0811.3894}},
  \href {http://dx.doi.org/10.1103/PhysRevLett.101.261104}
  {\path{doi:10.1103/PhysRevLett.101.261104}}.

\bibitem{2009arXiv0905.0105H}
{H.E.S.S. Collaboration: F. Aharonian}, ArXiv e-prints\href
  {http://arxiv.org/abs/0905.0105} {\path{arXiv:0905.0105}}.

\end{thebibliography}

\end{document}